\documentclass[prb,aps,amsmath,amssymb,reprint]{revtex4-1}
\usepackage{graphicx}
\usepackage{dcolumn}
\usepackage{bm}
\newcommand{\g}{$\gamma$}

\begin{document}
\title{Surfactant enhanced antiferromagnetic coupling in magnetron sputtered Cu/Co multilayers: A neutron reflectivity study}
\author{S. M. Amir, Mukul Gupta}
\email[]{mgupta@csr.res.in}
\author{Ajay Gupta}
\affiliation{UGC-DAE Consortium for Scientific Research,
\\University Campus, Khandwa Road, Indore-452 001,India}
\author{J. Stahn}
\affiliation{Laboratory for Neutron Scattering, Paul Scherrer
Institut, CH-5232 Villigen PSI, Switzerland}

\date{\today}

\begin{abstract}

In this work we studied Cu/Co multilayers prepared using
dc-magnetron sputtering technique with Ag surfactant. It was found
that Ag balances the difference in the surface free energy of Cu
and Co and this results in removing the asymmetry in the interface
roughness of Cu-on-Co and Co-on-Cu interfaces. As the interfaces
become symmetric, we observe a significant enhancement in
antiferromagnetic coupling and magneto resistance. Further, a
correlation of spin-dependent scattering with the interface
roughness is brought by comparing Cu/Co multilayer prepared using
different deposition methods. It was found that as interface
roughness increases spin-dependent scattering decreases.
\end{abstract}


\maketitle

\section{Introduction}

Magnetic layers separated by a non-magnetic spacer layer are
well-known to exhibit the interlayer exchange coupling (IEC)
between the magnetic
layers.~\cite{Baibich:PRL88,Parkin_PRL1990,Parkin_PRL1991} In
particularly, Cu/Co and Fe/Cr multilayers show the giant
magnetoresistanec (GMR) and an oscillatory exchange coupling with
a variation in the thickness of the Cu (Cr) spacer layer. Other
than the thickness of the spacer layer, intermixing at the
interface (due to chemical diffusion) and interface roughness
($\mathrm{\sigma_{int}}$) strongly influence the
GMR.~\cite{Hall:PRB93,Fullerton:PRL92} It is known that the GMR
basically originates from the antiferromagnetic coupling (AFC)
between magnetic layers and spin-dependent scattering of electrons
taking place within a layer (bulk scattering) and at the
interfaces.~\cite{Barnas:PRB1996,Hall:PRB93,Levi:PRL90} At an
interface there is a change in the electronic band structure
giving rise to asymmetric spin-dependent electron scattering. As
such it is rather difficult to separate the contribution of bulk
and interface scattering, some experimental results evidence that
scattering is larger at the interface as compared to
bulk.~\cite{Parkin_PRL1993,parkin:APL1992,baumgart:JAP91,AGupta:JPSJ2000}
In this context the role of $\mathrm{\sigma_{int}}$ is immensely
important to control GMR.

Experimental results reveal that Fe/Cr and Cu/Co multilayers show
an opposite-type behavior of GMR with the interface roughness. In
Fe/Cr multilayers, it was found that the electron scattering at
the interface is spin-dependent, irrespective of the deposition
method (e.g. sputtering or
evaporation).~\cite{Petroff:JMMM91,Fullerton:PRL92} However, in
Cu/Co multilayers results are contradictory. In MBE grown Cu/Co
multilayers Hall $\emph{et al.}$ concluded the electron scattering
at the interface is
spin-$independent$~\cite{Hall:JMMM1992,Hall:JPCM1992}, whereas in
sputtered multilayers Parkin $\emph{el al.}$~\cite{Parkin_PRL1993}
found that the electron scattering at the interface is
spin-$dependent$ similar to Fe/Cr multilayers. Such contradictory
observations of GMR in MBE grown and sputter deposited Cu/Co
multilayer are still debated. In the present work we compare Cu/Co
multilayers having different interface roughness and find a strong
dependence of electronic scattering on the interface roughness.

In a multilayer, the interface structure depends on several
parameters of which adatom energies (E$\mathrm{_a}$, kinetic
energy of the atoms being deposited on a substrate), deposition
rate (R$\mathrm{_d}$) and surface free energy (\g) of the elements
play a very important role on the interface
structure.~\cite{egelhoff_JVT1989,Copel_PRL89,Zhang_PRL94} For
example, \g \, determines the type of growth that takes place
while E$\mathrm{_a}$ and R$\mathrm{_d}$ are characteristic of a
deposition process. An element with low \g \, will wet the surface
of elements with high \g \, and make a smooth interface, whereas
high \g \,element agglomerates over the low \g\, element surfaces
and make a rough interface.~\cite{colin:Nature89} This leads to
asymmetric interfaces in a multilayer structure. In particularly
in Cu/Co multilayers, this difference in \g\, severally affects
the growth of multilayer as the average value of \g \, for
polycrystalline Cu and Co are 1.8\,J/m$^2$ and 2.55\,J/m$^2$,
respectively.~\cite{Foiles_PRB86,Tyson_SS77} This results in rough
Co-on-Cu and smooth Cu-on-Co interfaces.~\cite{Timothy:JAP96} It
has been shown experimentally that the growth mode of Cu/Co
multilayer can be altered by adding a third element known as
surfactant.~\cite{Camarero_PRL96,Chopra_PRB97,Ling:PRB2000,Gomez:PRB01,Amir:JPCM2011,Amir:JAC12}
Various types of surfactants e.g. Ag (\g=1.2 J/m$^2$), Sn(\g=0.65
J/m$^2$), Pb (\g=0.6 J/m$^2$) etc. (\g~ values for polycrystalline
case) have been used in different types of
multilayers.~\cite{Copel:PRB1990,Eaglesham:PRL1993,Oppo:PRL1993,Tromp:PRL1992,MG:APL2011}
It has been demonstrated that surfactant floats at the surface
balancing the \g\, of the elements, suppress surface diffusion and
prevents island
formation.~\cite{Copel_PRL89,Hoegen_PRL91,SiGe_APL05} To act as a
good surfactant an element must have relatively smaller \g\, and
larger volume so that its incorporation can be avoided.

Other aspects that influence the interface structure are
R$\mathrm{_d}$ and E$\mathrm{_a}$. For deposition of a particular
multilayer, although the \g\, will not vary, R$\mathrm{_d}$ and
E$\mathrm{_a}$ may change with the choice of deposition method.
Generally in magnetron sputtering process both R$\mathrm{_d}$ and
E$\mathrm{_a}$ ($\sim$10-20\,eV) are large, in thermal evaporation
process both R$\mathrm{_d}$ and E$\mathrm{_a}$ are low with
E$\mathrm{_a}$ being about two orders of magnitude smaller and
R$\mathrm{_d}$ typically an order of magnitude smaller. In ion
beam sputter (IBS) deposition E$\mathrm{_a}$ is large and tunable,
R$\mathrm{_d}$ is typically comparable to e-beam methods.
Therefore, by depositing samples using these different deposition
methods with a surfactant, the $\mathrm{\sigma_{int}}$ can be
varied precisely.

In our recent works the effect of Ag surfactant on the
$\mathrm{\sigma_{int}}$ was studied in Cu/Co multilayers prepared
using IBS~\cite{Amir:JPCM2011,MG_CuCo} and e-beam
evaporation.~\cite{Amir:JAC12} Although the values of
$\mathrm{\sigma_{int}}$ in these deposition processes were
different ($\mathrm{\sigma_{int}}$ = 0.1\, nm in IBS and 1\,nm in
e-beam), it was found that in both cases Ag surfactant helps in
removing the asymmetry of $\mathrm{\sigma_{int}}$ caused due to
different \g\, of Cu and Co. In the present work we apply
dc-magnetron sputtering (dc-MS) technique to study the effect of
Ag surfactant in Cu/Co multilayers and found that Ag surfactant
yield symmetric Cu/Co and Co/Cu interfaces. Here the value of
$\mathrm{\sigma_{int}}$ was found to be in between that of IBS and
e-beam. A correlation of $\mathrm{\sigma_{int}}$ with MR
normalized with antiferromagnetic fraction (AFF) clearly shows
that spin-dependent electron scattering at the interface of Co/Cu
multilayers decreases with increasing $\mathrm{\sigma_{int}}$. The
obtained results are presented and discussed in the following
sections.

\section{Experimental Details}

We deposited Cu/Co multilayers on silicon substrates using dc-MS
technique with following structure:
Cu\,(10\,nm)/$x$/[Cu\,(3\,nm)/Co\,(2\,nm)]$_{10}$, with $x$ = one
monolayer of Ag or 0 (reference sample without Ag surfactant).
With a base pressure of $\sim$$1\cdot10^{-7}$\,mbar, the
deposition was  carried out at $2\cdot10^{-3}$ mbar pressure using
5\,sccm Ar gas for sputtering. Circular targets of (75\,mm
diameter) of pure Cu, Co and Ag (purity 99.999\%) were sputtered
at power of 50\,W. All targets were cleaned by pre-sputtering for
about 10\,minutes. The deposition rates obtained were
31.3\,nm/min, 13.4\,nm/min and 40\,nm/min for Cu, Co and Ag,
respectively. Since Ag surfactant should be about a monolayer
thick, a 10\,mm slit was placed below Ag target which reduced Ag
deposition rate to about 5\,nm/min. The substrates were oscillated
linearly with respect to a central position of a target for better
uniformity of deposition area.

As-deposited Cu/Co multilayers were studied using unpolarized
neutron reflectivity (NR) and polarized neutron reflectivity
(PNR). These measurements were performed at the AMOR reflectometer
at Swiss Spallation Neutron Source (SINQ) at PSI,
Switzerland~\cite{Gupta_PJP04} in the time of flight geometry
using neutrons of wavelength 0.15nm$<$$\lambda$$<$1.3 nm. The
polarization efficiency of FeCoV/Ti:N supermirror polarizer was
about 97\%. X-ray diffraction (XRD) measurements were done in
$\theta$-2$\theta$ geometry using x-rays of wavelength 1.54\,\AA\,
in standard diffractometer (Bruker D8 Advance). The x-rays were
detected using a fast counting detector based on Silicon strip
technology (Bruker LynxEye detector). Magnetoresistance (MR)
measurements were performed using four point probe method at room
temperature. During the MR measurement direction of the magnetic
field was along the direction of the current flowing in the
sample. Magnetization (M) versus magnetic field (H) hysteresis
loop of the samples were recorded using superconducting quantum
interference device (SQUID)-vibrating sample magnetometer (SVSM;
Quantum Design Inc., USA).

\section{Results}

\subsection{X-ray diffraction measurements}

\begin{figure}
\begin{center}
\includegraphics[width=70mm,height=70mm]{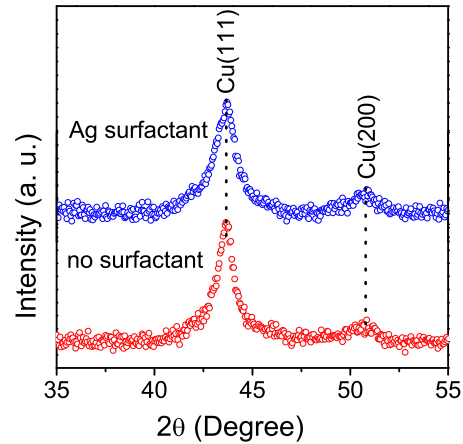}
\vspace{-5mm} \caption{\label{fig:1} XRD pattern of Cu/Co
multilayer samples prepared with and without Ag surfactant using
dc-MS method.\vspace{-1mm}}
\end{center}
\end{figure}

Fig.~\ref{fig:1} shows the XRD pattern of Cu/Co multilayers
prepared with and without Ag surfactant. Bragg peaks corresponding
to Cu (111) and Cu (200) reflections can be seen around
2$\theta$=44.5\,deg. and 50.6\,deg., respectively. The average
grain size can be estimated using Scherrer formula and found to be
around 6\,nm in both samples. It appears that the grain growth is
not significantly affected due to the presence of Ag surfactant in
the Cu/Co multilayers. This is in agreement with Cu/Co multilayers
prepared using Ag surfactant with other deposition
methods.~\cite{Amir:JPCM2011,Amir:JAC12}

\subsection{Neutron reflectivity measurements}

\begin{figure}
\begin{center}
\includegraphics[width=90mm,height=70mm]{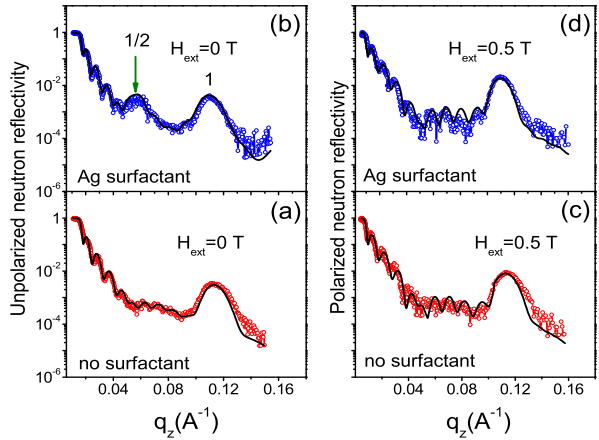}
\vspace{-5mm} \caption{\label{fig:2} Unpolarized and spin-down
($\mathcal{R^-}$) neutron reflectivity of Cu/Co multilayer samples
prepared without Ag surfactant (a,c) and with Ag surfactant (b,d)
using dc-MS method. Scattered points are experimental data and
solid lines are fit to them. \vspace{-1mm}}
\end{center}
\end{figure}

NR and PNR measurements were performed to probe the interface
structure in Cu/Co multilayer samples. Fig.~\ref{fig:2}(a,b) shows
the NR pattern of Cu/Co multilayers prepared without and with Ag
surfactant. The NR measurements were performed without exposing
the samples to any magnetic field. The reflectivity pattern of
samples shows first order nuclear Bragg peak around momentum
transfer vector q$_z$=0.11\,\AA$^{-1}$
($q_z=4\pi\,\mathrm{sin}\theta/\lambda $, where $\theta$ is
incidence angle and $\lambda$ is wavelength of neutrons). This
correspond to a bilayer period of about 5.6\,nm. The sample
prepared with Ag surfactant shows in addition a half-order peak
around q$_z$=0.055\,\AA$^{-1}$. This half-order Bragg peak
originates as Co layers gets antiferromagnetically coupled and
give rise to additional magnetic superstructure at twice the
period of nuclear structure. No such magnetic peak could be
observed in the NR pattern of sample where Ag surfactant was not
used. To understand the observed results, the interface structure
of the samples prepared with and without Ag surfactant should be
analyzed. The fitting of the NR pattern provides detailed
information about the Cu/Co and Co/Cu interfaces. The NR patterns
were fitted using Parratt's formalism ~\cite{Parratt32}, the
fitting parameters are given in table~\ref{tab:table1}. As can be
seen from the table, the interface roughness of Co-on-Cu and
Cu-on-Co interfaces becomes almost equal when Ag surfactant is
added.

\begin{table*}
\caption{\label{tab:table1} Cu-on-Co and Co-on-Cu interface
roughness ($\mathrm{\sigma_{int}}$) of the Cu/Co multilayer
samples prepared with and without Ag surfactant with nominal
structure of Si(100)/Cu(10\,nm)/[Cu(3\,nm)/Co(2\,nm)]$_{10}$ using
dc-Magnetron sputtering (dc-MS), ion beam sputtering (IBS) and
e-beam evaporation techniques.\\}
\begin{ruledtabular}
\begin{tabular}{c|c|c|c|c}
Deposition Method&$\mathrm{\sigma_{int}}(\mathrm{nm})$&No
surfactant & Ag surfactant&Ref.\\ \hline
dc-MS&& &\\
&$\sigma$$_{\mathrm{[Co-on-Cu]}}$ & 0.74$\pm0.05$  & 0.4$\pm0.05$&Present work\\
&$\sigma$$_{\mathrm{[Cu-on-Co]}}$ & 0.32$\pm0.05$  & 0.4$\pm0.05$&\\

IBS&& &\\
&$\sigma$$_{\mathrm{[Co-on-Cu]}}$ & 0.36$\pm0.05$  & 0.11$\pm0.05$&~\cite{Amir:JPCM2011}\\
&$\sigma$$_{\mathrm{[Cu-on-Co]}}$ & 0.18$\pm0.05$  & 0.10$\pm0.05$&\\

e-beam&& &\\
&$\sigma$$_{\mathrm{[Co-on-Cu]}}$ & 1.6$\pm0.1$  & 1.1$\pm0.05$&~\cite{Amir:JAC12}\\
&$\sigma$$_{\mathrm{[Cu-on-Co]}}$ & 1.0$\pm0.1$  & 1.0$\pm0.05$&\\

\end{tabular}
\end{ruledtabular}
\end{table*}

\begin{figure}
\begin{center}
\includegraphics[width=70mm,height=70mm]{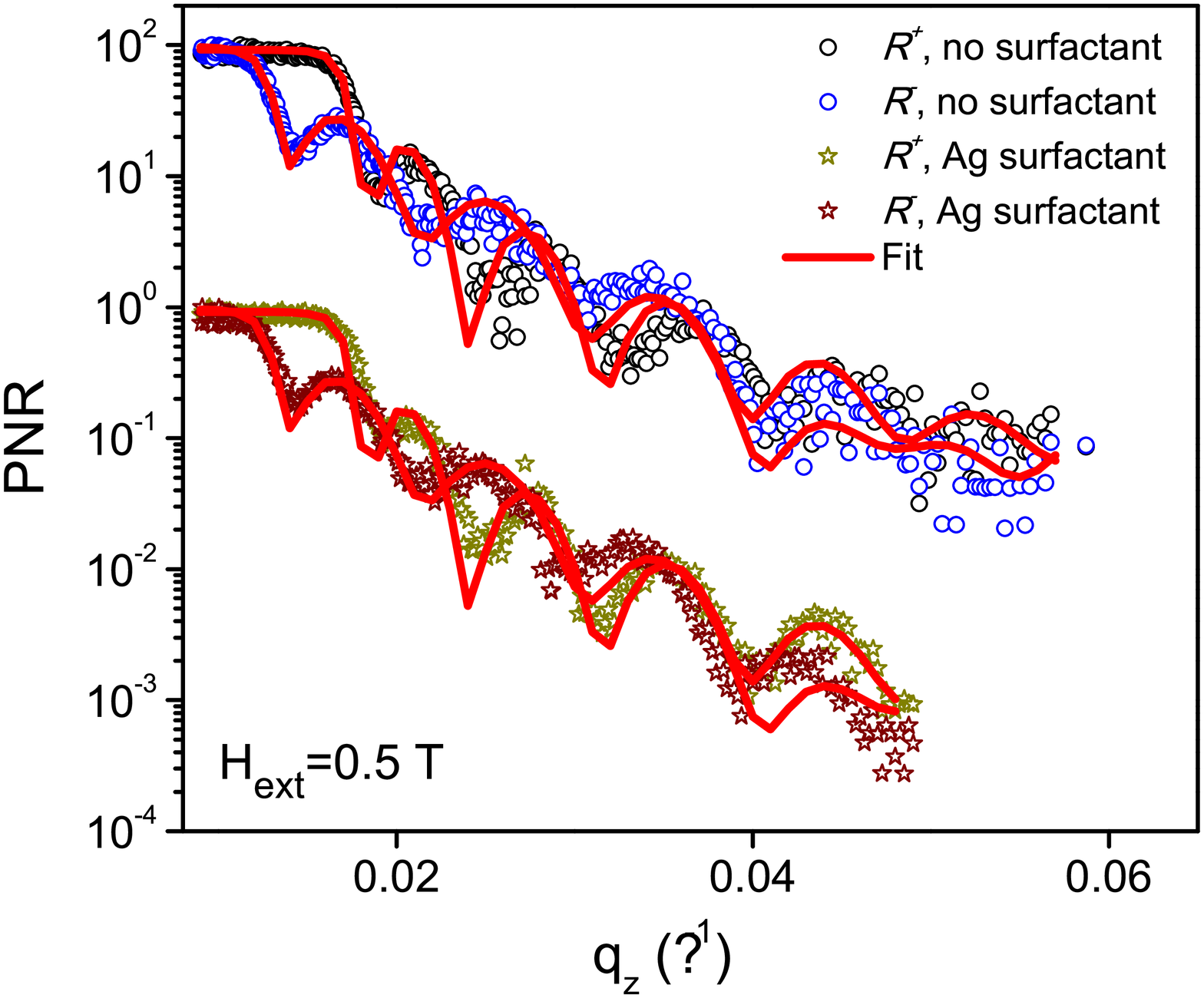}\vspace{-5mm}
\caption{\label{fig:3} PNR pattern of the samples prepared without
surfactant and with Ag surfactant using dc-MS method. The PNR
pattern of samples prepared without surfactant have been shifted
on y-scale for clarity.}
\end{center}
\end{figure}

Further to confirm that this half-order peak appearing in the Ag
surfactant sample indeed has magnetic origins, we performed PNR
measurement by applying an external magnetic field of 0.5\,T
parallel to the sample. This magnetic field strength is sufficient
to saturate the samples magnetically. Fig.~\ref{fig:3} shows the
PNR data of samples around the critical angle. PNR gives a precise
information about the absolute magnetic moment per atom in a
magnetic sample and this is independent of substrate magnetism
(often paramagnetic or diamagnetic) or sample volume as in case of
bulk magnetization techniques e.g. SQUID, VSM or extraction
methods.~\cite{Mukul:NJP2008} This makes PNR a unique technique to
measure absolute magnetic moment in magnetic samples. The PNR data
of the samples were fitted using a computer
software~\cite{Simulreflec} and it reveals that the Co magnetic
moment in both samples is about 1.7\,$\mathrm{\mu_B}$ per atom,
which is close to bulk Co magnetic moment. This indicates that Ag
surfactant has no influence on the magnetic moment of these
samples. Fig.~\ref{fig:2} (c,d) shows the spin down
($\mathcal{R^-}$) PNR pattern for both samples to higher q$_z$. It
may be noted that the contrast with polarized neutrons for Cu and
Co is larger for spin-down neutrons. For neutrons the nuclear
scattering length density (SLD) for Cu and Co is
$6.55\times10^{-6}$\, \AA$^{-2}$ and $2.26\times10^{-6}$\,
\AA$^{-2}$, respectively whereas magnetic SLD of Co is
$4.22\times10^{-6}$\, \AA$^{-2}$. For spin-up neutrons the
contrast between Cu and Co becomes $6.55\times10^{-6}$\ \AA$^{-2}$
and $6.48\times10^{-6}$\ \AA$^{-2}$ whereas for spin-down neutrons
it is $6.55\times10^{-6}$\ and $-1.96\times10^{-6}$ \AA$^{-2}$,
obviously a larger contrast for spin-down neutrons permits to
probe the Cu/Co multilayers more accurately. As can be seen from
Fig.~\ref{fig:2} (d) the half order peak has disappeared
completely after applying a magnetic filed of 0.5\,T, confirming
that the half order peak was indeed arisen due to
antiferromagnetic coupling of alternate Co layers. A detailed
fitting of PNR data reveals that the $\mathrm{\sigma_{int}}$ in
the multilayers are similar to those obtained with NR data. This
confirms that the structural roughness of Cu-on-Co and Co-on-Cu
interfaces become equal when Ag surfactant was used.

\subsection{Magnetotransport measurements}

\begin{figure}
\begin{center}
\includegraphics[width=70mm,height=70mm]{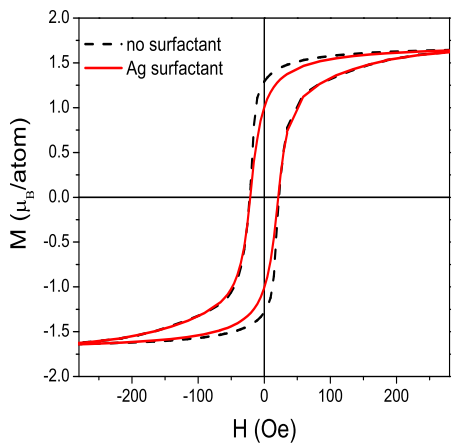}\vspace{-5mm}
\caption{\label{fig:4} Magnetizaton (M) versus magnetic field (H)
of samples prepared with and without Ag surfactant using dc-MS
method.}
\end{center}
\end{figure}

\begin{figure}
\begin{center}
\includegraphics[width=70mm,height=70mm]{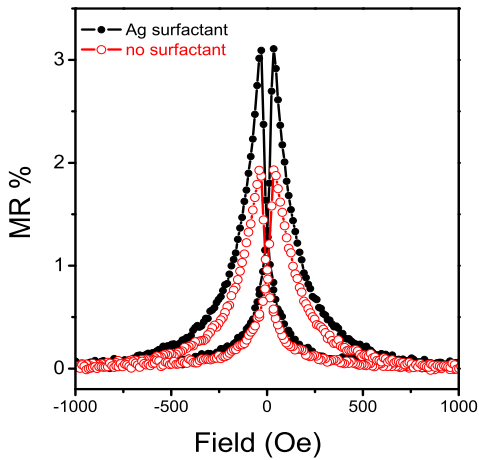}\vspace{-5mm}
\caption{\label{fig:5} MR of samples prepared with and without Ag
surfactant using dc-MS method.}
\end{center}
\end{figure}

Fig.~\ref{fig:4} shows the M-H hysteresis loop of the samples
prepared with and without Ag surfactant. As can be seen, there is
no appreciable change in the coercivity and saturation
magnetization (M$_s$, also confirmed by PNR measurements) of the
samples but the remanent magnetization (M$_{r}$) is significantly
different. As discussed before, the addition of Ag surfactant
enhances the AFC in Cu/Co multilayer (see Fig.~\ref{fig:2} b). The
AF coupled region can be quantified as AFF = 1-
M$_{r}$/M$_{s}$.~\cite{Chen:JPCM08,Schad:PRB99,Amir:JPCM2011} We
find that the AFF for the sample prepared without Ag surfactant is
about 18.8\% and increases to about 39.2\% when Ag surfactant is
used. This enhancement of AFF for the sample prepared with Ag
surfactant re-confirms our observation from NR measurements, where
the intensity of half-order peak gets enhanced when Ag surfactant
is added.

For studying the effect of Ag surfactant, on the MR of the samples
we performed MR measurements using the four-probe method.
Fig.~\ref{fig:5} shows the MR of the samples prepared with and
without Ag surfactant. Here MR is defined as MR= (R$_0$ -
R$_{s}$)/R$_0$, where R$_0$ is the resistance in the absence of
magnetic field and R$_s$ is the resistance under the saturation
magnetic field. The MR of the samples prepared with and without Ag
surfactant are about 3.1$\%$ and 1.9$\%$ respectively. It may be
noticed that there is significant increase in MR for the sample
prepared using Ag surfactant.

\section{Discussion}

We observe that the addition of Ag surfactant helps in removing
the asymmetry in the $\mathrm{\sigma_{int}}$ of Cu-on-Co and
Co-on-Cu interfaces independent of deposition method (see
table~\ref{tab:table1}). Similar observation were also made by
several other groups in Cu/Co
multilayers.~\cite{Egelhoff_JAP1996,Tolkes_PRL98,Camarero_PRL94,Cho:PRB1998,Muller:JPCM2001}
However, an enhancement in MR and AFC as observed in NR
measurement is remarkable as: (i) the thickness of Cu layer
corresponds to third and weakest maxima of the oscillatory
exchange coupling (ii) the number of bilayer repetitions taken
were only 10.

\begin{table}
\caption{\label{tab:table2} Average value of adatom energy
(E$\mathrm{_a}$) and deposition rate (R$\mathrm{_d}$) for Cu/Co
multilayer samples prepared using dc-Magnetron sputtering (dc-MS),
ion beam sputtering (IBS) and e-beam evaporation techniques.\\}
\begin{ruledtabular}
\begin{tabular}{c|c|c}
Deposition Method&E$\mathrm{_a}$ (eV)&R$\mathrm{_d}$ (nm/min) \\
\hline

dc-MS&$\sim$18&$\sim$30 \\
IBS&$\sim$22&$\sim$4.5\\
e-beam&$\sim$0.2&$\sim$1.2\\

\end{tabular}
\end{ruledtabular}
\end{table}

More insight about spin-dependent scattering can be obtained by
comparing Cu/Co multilayer with different $\mathrm{\sigma_{int}}$.
Table~\ref{tab:table1} shows the value of $\mathrm{\sigma_{int}}$
observed in Cu/Co multilayers prepared with dc-MS, IBS and e-beam
evaporation techniques. As mentioned before, these deposition
techniques are different either in term of E$\mathrm{_a}$ or
R$\mathrm{_d}$ which leads to the difference in the
$\mathrm{\sigma_{int}}$ in Cu/Co multilayers. The average value of
E$\mathrm{_a}$ and R$\mathrm{_d}$ involved in different deposition
techniques are given in table~\ref{tab:table2}(calculated using
SRIM simulations~\cite{SRIM}). In case of e-beam evaporation
technique E$\mathrm{_a}$ directly translates to melting
temperature of Cu, Co and Ag whereas in case of IBS and dc-MS,
E$\mathrm{_a}$ varies with Ar$^+$ energy. As can be seen from
table~\ref{tab:table2} that E$\mathrm{_a}$ values for IBS and
dc-MS are not significantly different but the R$\mathrm{_d}$ is an
order of magnitude larger than the IBS technique. With the help of
known E$\mathrm{_a}$ and deposition rates, the growth mode of
Cu/Co multilayers can be estimated. In case of e-beam island type
growth is well-known and roughness keeps on increasing with the
thickness.~\cite{Zhao:COL2010} For IBS very large values of
E$\mathrm{_a}$ and slow deposition rate result in almost
layer-by-layer type growth~\cite{Mukul:ASS2003} and in dc-MS high
E$\mathrm{_a}$ and R$\mathrm{_d}$ essentially results in
layer-by-layer growth followed by island type growth. This
explains different values of $\mathrm{\sigma_{int}}$ obtained in
these samples.

\begin{figure}
\begin{center}
\includegraphics[width=80mm,height=75mm]{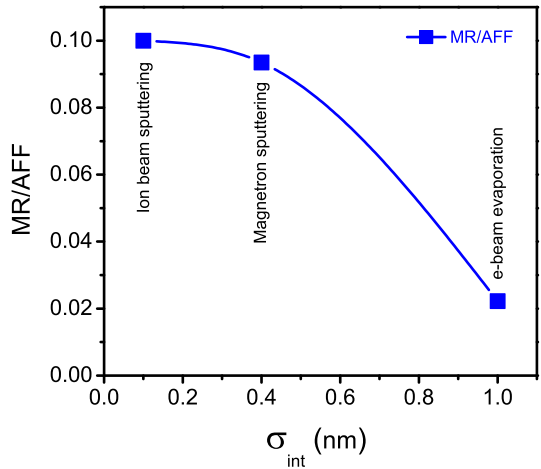}\vspace{-5mm}
\caption{\label{fig:6} Variation of normalized magnetoresistance
to antiferromagnetic coupling fraction (MR/AFF) with interface
roughness. The scattered points are measured data and solid lines
are guide to the eye.}
\end{center}
\end{figure}

We can now compare the spin-dependent scattering with
$\mathrm{\sigma_{int}}$ in samples prepared using different
methods with Ag surfactant. In magnetic multilayers, MR originates
due to the competitive process of spin dependent scattering and
AFC and MR normalized with AFF exhibits only the spin dependent
scattering.~\cite{Amir:JPCM2011,Colino:PRB1996,kopcewicz:JAP2003}
A plot of MR/AFF in fig.~\ref{fig:6} shows a decreases in it as
$\mathrm{\sigma_{int}}$ increases. This is a clear evidence
showing that the spin-dependent scattering decreases with an
increase in interface roughness in Cu/Co multilayers. This
essentially implies significance of scattering from the interfaces
in Cu/Co multilayers.

\section{Conclusions}

We observe Ag surfactant leads to symmetric interfaces in Cu/Co
multilayers. Symmetric interface roughness enhances the AFC in
Cu/Co multilayer which give rise to enhanced MR in the sample. A
comparison of the Cu/Co multilayers prepared using different
techniques - ion beam sputtering, e-beam evaporation and
dc-magnetron sputtering reveal that spin-dependent scattering at
the interface decreases with increase in $\mathrm{\sigma_{int}}$.
Observed results help in establishing the fact that spin-dependant
electron scattering is significant in Cu/Co multilayers.

\acknowledgements {This work is partially based on experiments
performed with neutron reflectometer AMOR at the Swiss spallation
neutron source SINQ, Paul Scherrer Institute, Villigen,
Switzerland. We acknowledge DST, Government of India for providing
financial support to carry out NR experiments under its schemes
`Utilisation of International Synchrotron Radiation and Neutron
Scattering facilities'. We are thankful to M. Horisberger and S.
Potdar for their help in sample preparation. We are also thankful
to Dr. R. J. Chaudhary for SVSM measurement. Continuous support
and encouragements received from Dr.\,P.\,Chaddah is gratefully
acknowledged.}

%
\end{document}